\documentstyle[12pt,epsf,aaspp4]{article}
\slugcomment{\shortstack[r]{KUNS-1558}}

\begin{document}

\title{The Sunyaev-Zel'dovich Effect by Cocoons of Radio Galaxies}
\author{Masako Yamada\altaffilmark{1} \and Naoshi Sugiyama}
\affil{Department of Physics,Kyoto University, Kyoto 606-8502,Japan}
\altaffiltext{1}{e-mail:masako@tap.scphys.kyoto-u.ac.jp}
\and

\author{Joseph Silk}
\affil{Departments of Astronomy and Physics, and Center for Particle
  Astrophysics, University of California, Berkeley, CA94720}

\begin{abstract}
We estimate the deformation of the cosmic microwave background
radiation by the hot region (``cocoon'') around a radio galaxy.
A simple model by  is adopted for cocoon evolution  while the
jet is on, and a model of evolution  is
constructed after the jet is off.  
It is found that at low redshift 
the phase after the jet is off is
longer than the lifetime of the jets. 
The Compton $y$ parameter generated by cocoons is calculated with a
Press-Schechter number density evolution. 
The resultant value of $y$ is of the same order as the
{\it COBE} constraint. 
The Sunyaev-Zel'dovich effect due to cocoons could therefore be a significant
foreground source of small angular scale  anisotropies in
 the cosmic microwave background
radiation. 
\end{abstract}


\section{Introduction}
Ionized gas deforms the spectrum of the cosmic microwave background
radiation (CMB) via inverse Compton scattering,  the resulting
spectral distortion being known as the
Sunyaev-Zel'dovich effect (Sunyaev \& Zel'dovich 1980; see for a recent
review Birkenshaw, 1998). 
The most popular source of the Sunyaev-Zel'dovich effect is  hot
X-ray emitting gas in a cluster of galaxies. 
The effects of the intracluster gas have been  estimated theoretically 
by many authors
 (see e.g.,  
Makino \& Suto 1993;Aghanim et al. 1997), and there exist observations of the
temperature decrement towards rich clusters of galaxies. 
Recently Aghanim et al.(1996) studied the effects of the partial
ionization of the intergalactic medium (IGM) by quasar UV radiation.
They showed that
the kinematical effect due to the peculiar motions of the ionized
bubbles is much larger than the thermal effect and can be
detected by future  generations of CMB observations. 
Natarajan \& Sigurdsson (1999) exmanied the Sunyaev-Zel'dovich effect in 
the region formed by the quasar outflows.
We examine here another  source of the Sunyaev-Zel'dovich effect which is 
associated with radio galaxies.

It has been suggested that a standard model for a typical
strong double radio
source comprises  twin  jets advancing into the intergalactic medium and
surrounded by a hot region `cocoon' that consists of jet-supplied matter
and shocked IGM (Scheuer 1974; Blandford \& Rees 1974). 
Begelman \& Cioffi (1989) and Nath (1995) constructed a model of
the evolution of a cocoon in which the cocoon is overpressured against the
IGM during its initial evolution. 
In their model, the expansion along the jet axis is determined by the
balance of the thrust of the jet and the ram pressure, and the thermal
pressure of the cocoon drives the expansion
along the direction perpendicular to the jet axis.   
It has been demonstrated that some numerical simulations agree with this
scenario (Loken et al. 1992; Cioffi \& Blondin 1992). 

The energy density of the cocoon is higher than that of the
intergalactic medium, and therefore it works as a local hot region that
induces the thermal Sunyaev-Zel'dovich effect  much as is observed for 
the hot intracluster gas in 
galaxy clusters. 
When we consider the lifetime of the jet($\sim
10^7-10^8$ year), which 
is considered to be shorter than the Hubble time at the
redshift where 
radio galaxies reside, cocoons may not play a significant role in  
the deformation of the CMB spectrum.
However, even after the jet energy supply  stops, cocoons
can remain hot; the cocoon expands due to its own thermal pressure, and
the thermal energy of the cocoon is consumed by expansion and radiative
cooling by the cocoon gas.
Thus the cocoon remains a possible source of the
Sunyaev-Zel'dovich effect for an effective lifetime 
that can be much longer than the jet lifetime.

We estimate the Sunyaev-Zel'dovich effect due to
 the cocoons by taking into account
their evolution both while the jet is on and after the jet is off.
We make use of the simple model of Nath (1995) 
for the evolution while the jet is on, and first estimate a rough
order of 
$y$ for a single cocoon.
We construct a model for the
evolution after the jet is off, 
and evaluate the average Compton $y$ parameter for  a uniform
distribution of cocoons. 
Comparing our result with the {\it COBE} constraint, we discuss the
possibility of detection of the effect of cocoons in the CMB
observations.

Our paper is organized as follows.
In \S2 we summarize our basic calculation and list the
parameters we need to estimate in order to apply   evolutionary models for the
cocoons.
In \S3 and \S4 we describe  models of cocoon evolution during the period while, and after, 
the jet is on.
In \S5 we  calculate the  Sunyaev-Zel'dovich effect and give our results.
Discussion and conclusions are given in \S6.

\clearpage

\section{Basics and Necessities}
First we represent the formal expression of the Compton $y$ parameter
and clarify what parameters we need to estimate.
For a single cocoon, 

\begin{eqnarray}
y_{\rm sc} &\equiv& \int\frac{kT}{m_ec^2}n_e\sigma_TdR, \nonumber \\
       &\simeq& \frac{P_e}{m_ec^2}\sigma_TV^{1/3},
\end{eqnarray}
where $\sigma_T$, $n_e$, $T$, $P_e$, $dR$ are the Thomson scattering
cross section,  
electron number density, electron temperature, pressure, 
and distance element along the line 
of sight, respectively.
In the second equation, we set the length
scale along the line of sight $R\simeq V^{1/3}$ as averaged for 
directions over a randomly oriented distribution of cocoons.
The total $y$ in a line of sight with a solid angle element $d\Omega$ is 
the integral of the product of $y_{\rm sc}$, the formation rate of
cocoons, the volume element $a^3r^2drd\Omega$, and the crossing
probability, as

\begin{eqnarray}
y &=& \int\int\int y_{\rm sc}(M,t-t_{\rm col})\frac{\partial n}{\partial
  t}(M,t_{\rm col})dt_{\rm col}a^3r^2drd\Omega\frac{\theta^2}{d\Omega}dM,
  \nonumber \\
  &\approx& \int\int y_{\rm sc}(M,z)n(M,z)a^3r^2dr \left( \frac{R}{R_A} \right)^2
  dM, \nonumber \\
  &\approx& \int\int \frac{P_eV}{m_ec^2}\sigma_Tn_(M,z)a^3r^2dr\frac{1}{R_A^2}dM.
\end{eqnarray}
Here, $a$ is the cosmological 
scale factor, $r$ is the comoving radial coordinate,
$R_A$ is the angular diameter distance, $\theta=R/R_A$,
$n$ is the number density of radio galaxies with mass $M$ at $t_{\rm
  coll}$, $\theta^2/d\Omega$ is the  probability for the line
of sight to traverse
 the cocoon, respectively.
The angular diameter distance $R_A$ is given by
\begin{equation}
R_A = a(z)r=\frac{1}{(1+z)}\frac{2c}{H_0}[1-(1+z)^{-1/2}].
\end{equation}
Strictly speaking, other than their mass scales, the epoch of the
collapse of objects depends on various factors such as the amplitude or
density distribution within each object. 
However, in the second and third equations, we assume for simplicity
that the 
objects which have the same mass
collapse at almost the same epoch, and replace $\partial n/\partial t\times
dt_{\rm col}$ with $n(M,z)$.
Thus in order to evaluate $y$, we only need the value of $P_eV$ for each
cocoon.
In \S 3 and 4 we examine 
the evolution of $P_eV$ within a cocoon with the jet on
and off. 

\clearpage
\section{Model 1: while the Jet is On}
In this section we consider an evolutionary 
model of a strong double radio source with jets (Scheuer, 1974;
Blandford \& Rees  
1974; Begelman \& Cioffi 1989; Nath 1995) and make a rough estimate of
$y$ for a single cocoon.
We follow the model proposed by Nath (1995), in which the jets are
enveloped in a hot cocoon consisting of shock-compressed jet matter and
intergalactic medium (IGM).
In this model, the jets are confined by the pressure of the cocoon, i.e., 
we do not take confinement by magnetic pressure into account.
Highly collimated, `light' (the density of the jet matter is
smaller than the ambient IGM) jets interact with the IGM and form bow
shocks.
At the shock, some of the electrons are accelerated to high energy and 
radiate synchrotron emission.
The thermalization efficiency is one of the unknown parameters in this
model; however, because in many lobes of radio galaxies, magnetic fields
have been 
observed, we expect that the effective mean free path of the particles
should be short and the thermalization efficiency should be high because
of the interaction with the magnetic field.
Therefore in this paper, the relativistic jet matter is assumed to be 
completely thermalized at the interaction interface with the IGM.
The thermalized jet matter forms a high temperature region around the
jet together with the shock-compressed intergalactic medium.
The shock-compressed intergalactic medium surrounds the
thermalized jet matter separated by a contact discontinuity 
(see Figure \ref{image}). 
At least initially, the thermal pressure of this hot region 
is higher than the IGM pressure.
When the hot region pressure is higher than that of the IGM, the hot region
also expands 
in the direction perpendicular to the jet axis and is accompanied by shocks.
We make use of the term `cocoon' for the hot region surrounded by the shock
surface.
As the cocoon expands the pressure decreases.
Nath (1995) estimated the equilibrium timescale, i.e., the time required
until the cocoon pressure decreases sufficiently with
time to be equal to the ambient pressure:

\begin{figure}[t]
\centerline{\epsfxsize=\textwidth \epsffile{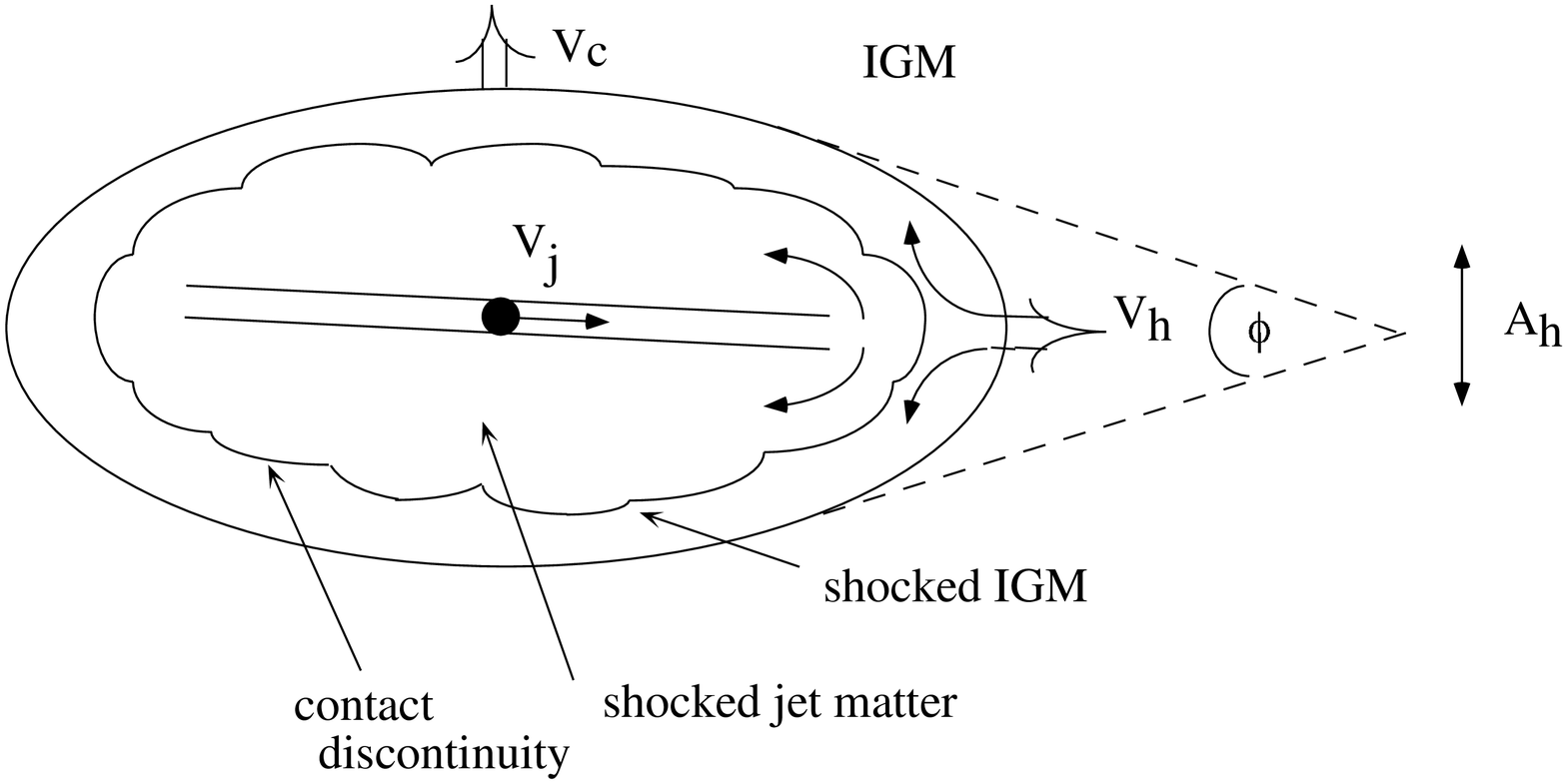}}
\caption{\protect\footnotesize
A schematic diagram of our model of the cocoon
  surrounding the jet while the jet is on. See text for details. 
\label{image}}
\end{figure}

\begin{equation}
t_{\rm eq}\sim 1.34\times10^{10}\left( \frac{T_a}{10^6{\rm K}} \right)
 ^{-1}\Omega_{\rm IGM}^{-1/4}h_{50}^{1/2}\left( \frac{L_j}{10^{45}{\rm
 ergs/sec}} \right) ^{1/4}\beta_j^{1/4}\left( \frac{A_h}{(30 {\rm
 kpc})^2} \right) ^{1/4}\left( \frac{\epsilon_v}{1/3} \right)
 ^{-1/2}{\rm yr},
\end{equation}
where $L_j$ is the kinetic power of the jet, $v_h$ is the shock advance
speed into the IGM, $\beta_j\equiv v_j/c$ is the jet speed, $A_h$ is the 
effective cross section of the bow shock, $\epsilon_v$ is the volume
factor (which depends on the shape of the cocoon), $\Omega_{\rm IGM}$ is 
the density of the IGM in unit of the critical density,
$H_0=50h_{50}$km/sec/Mpc, respectively, and the suffix $a$ denotes
IGM value. 
It is found with observed values
of Cygnus A that $t_{\rm eq}$ is longer than the jet lifetime as long as 
$T_a\lesssim 10^8$K: i.e.,
cocoons remain overpressured while the jet is on.

Hereafter we consider two collimated steady jets advancing into the
 ambient IGM.
The thermalized jet matter and the shock-compressed IGM matter form a
 cocoon around the jets and the cocoon expands with shocks advancing in
 directions both parallel and perpendicular to the jet axis.
Below we will derive some relations between the various variables in order
 to obtain $P_cV$ during this stage of evolution.

At the shock, the jet thrust is balanced by the ram pressure of the IGM at 
the shock that works on the effective area $A_h$,

\begin{equation}
\frac{L_j}{v_j}=A_h\rho_a v_h^2.
\end{equation}
The kinetic power $L_j$ is written as follows with jet speed $v_j$.
From observations and theory of the superluminal motions of the
quasars, $v_j$ is considered to be approximately equal to the 
speed of light. 
Therefore,

\begin{eqnarray}
L_j &=& \frac{1}{2}\rho_jv_j^3A_j,  \\
v_j &=& c.
\end{eqnarray}
Here $A_j$ is the cross section of the jet.
At the interaction of the jet and IGM, a bow shock is formed.
Following Nath (1995), we approximate the bow shock as two oblique
shocks.
Then the jump condition at the oblique shocks is
\begin{equation}
\frac{P_c}{P_a}=\frac{5(v_h/c_a)^2\sin^2\phi-1}{4},
\end{equation}
where $\phi$ is the angle between two oblique shocks, and $c_a$ is the sound
velocity of the ambient medium.

As we mentioned above, because the cocoon is overpressured against the
IGM, the shock also advances into the IGM in the direction
perpendicular to the jet.
The jump condition there is
\begin{equation}
P_c\approx\rho_a v_c^2,   \label{vc}
\end{equation}
where $v_c=dr_c/dt$ is the shock advance speed.

Finally, we assume that the internal energy within the cocoon is
supplied by the jet matter, and that jet kinetic energy is transformed
into the thermal energy of the cocoon gas and the kinetic energy to
expand the cocoon.
We also assumed the fraction of input energy that is converted to other
forms of energy (magnetic fields or high energy cosmic rays: see
discussion in \S6) can be neglected.
From strong shock conditions, $v_c$ is the same order as the random
velocity of the cocoon gas, and therefore the fraction of the thermal energy
to the total input energy $U_{\rm IGM}/E_{\rm input}$ is 
${\cal O}(1)$, 
\begin{eqnarray}
\frac{1}{\gamma-1}P_cV &\approx& L_jt,  \label{lj}\\
V &=& \epsilon _v r_c^2v_ht,
\end{eqnarray}
where $\gamma$ is the adiabatic index and $t$ is the time elapsed since the
jet is on. 
The final equation comes from
the assumption that $v_h$ is constant $l_h\approx v_ht$.
Numerical simulations indicate that $v_h$ decreases and $A_h$ increases
with time (Loken et al. 1992; Cioffi \& Blondin, 1992).
However, the time-dependence is relatively small ($l_h\propto t^{\alpha}$ 
with $\alpha\sim 0.8$ in their simulations), and therefore we assume
$v_h$ is constant for the estimate of $y_{\rm sc}$.
Note that for the estimate of total $y$, as described in \S2, we only
need $P_cV$, and therefore do not need the detailed evolutionary scenario
(see eq.[\ref{lj}]).

In order to estimate the Compton $y$ parameter
value, we need the number density $n(M,z)$ and $P_cV$ of objects with 
mass $M$ and redshift $z$.
As we show in \S 5, we use the Press-Schechter formalism for $n(M,z)$.
Therefore we need to write the product of the pressure of the cocoon 
$P_c$ and $V$ as function of $t$, mass of the host galaxy $M$, and 
the parameters of the ambient medium.
We make additional assumptions about the jet kinetic luminosity. 
We assume that $L_j$ is equal to the 
Eddington luminosity of the black hole $L_{\rm
  edd}=4\pi cGM_{\rm BH}/\kappa_e\approx 1.25\times 10^{38}(M_{\rm BH}/M_{\sun})$
ergs/sec.
For the mass of the black hole some recent observations showed that 
there is a rough proportionality between
the mass of the black hole and the host galaxy, $M_{\rm BH}=0.002M$
(Magorrian et al. 1998; Ford et al. 1997; van der Marel 1997).
Using these assumptions we can write $L_j$ in terms of the mass of the
host galaxy.
Integrating equation (\ref{vc}), we obtain the relation 

\begin{equation}
r_c^2=2\left( \frac{\gamma-1}{\epsilon_v}\frac{L_j}{\rho_av_h} \right)
^{1/2}t  ~ .
\end{equation}

Using the above equations we obtain the 
following expressions for $P_c$ and $V$,

\begin{eqnarray}
P_c &=& \left[ \frac{5}{8}\left\{ \frac{L_j(\gamma-1)}{\epsilon_vt^2}\right\}
    ^2\frac{\rho_a^2P_a}{c_a^2}\sin^2\phi\right] ^{1/5}, \label{Pc} \\
V &=& \frac{(\gamma-1)}{P_c}L_jt ~ . \label{V}
\end{eqnarray}  

According to the usual discussions of  
synchrotron 
spectral aging, the minimum age of 
the jet is typically about $10^{6-7}$ years (Liu,
Pooley, \& Riley 1992).
So if we take cosmological values for the ambient medium with
$\Omega_bh^2=0.01$, 

\begin{eqnarray}
P_c &\sim& 2.13\times 10^{-12}\left( \frac{M}{10^{12}M_{\sun}}
\right)^{2/5} \left( \frac{f_{\rm BH}}{0.002} \right)^{2/5} \nonumber \\
 &\times&\left( 
\frac{t_{\rm life}}{3\times 10^7 {\rm years} } \right)^{-4/5} (1+z)^{9/5} \left
( \frac{\sin\phi}{\sin 20^{\circ}}\right)^{2/5}  {\rm ergs/cm^3}, \\
R &\sim& V^{1/3}, \nonumber  \\
  &\sim& 1.34\left( \frac{M}{10^{12}M_{\sun}} \right)^{1/5} \left
      ( \frac{f_{\rm BH}}{0.002} \right)^{1/5} \nonumber \\
  &\times& \left( \frac{t_{\rm life}}{3\times 10^7 {\rm years}}
  \right)^{3/5} (1+z)^{-3/5}\left( \frac{\sin\phi}{\sin 20^{\circ}}\right)^{-2/15} {\rm Mpc},
\end{eqnarray}
and $y$,

\begin{eqnarray}
y_{\rm sc} &=& \int\frac{kT_e}{m_ec^2}n_e\sigma_T dR,
 \nonumber \\
 &\sim& \frac{P_c}{m_ec^2}\sigma_TV^{1/3}, \nonumber \\
 &\sim& 7.14\times 10^{-5} \left( \frac{M}{10^{12}M_{\sun}} \right)^{3/5} \left
      ( \frac{f_{\rm BH}}{0.002} \right)^{3/5} \nonumber \\
  &\times& \left( \frac{t_{\rm life}}{3\times 10^7 {\rm years}}
  \right)^{-1/5} (1+z)^{6/5}\left( \frac{\sin\phi}{\sin 20^{\circ}}\right)^{4/15}.
\end{eqnarray}
Thus we expect that the deformation of CMB caused by cocoons is possibly 
relevant to observations.
\clearpage
\section{Model 2: after the Jet is Off}
After the relatively short lifetime $\sim 10^7$ years, the jet turns off.
However, after the jet is off,
cocoons may retain a high temperature and remain  a source of the 
Sunyaev-Zel'dovich effect for a long time.
A cocoon cools radiatively and expands due to its own thermal
overpressure.
We need to estimate the effective lifetime of the cocoon with the jets off, 
as a `relic' of the radio
galaxy.

A cocoon can be considered as an energy input into the intergalactic medium.
Thus the evolution of a cocoon after the jet is off resembles the
evolution of a supernova remnant in the interstellar medium.
We may therefore adopt the analogy with the evolution of a supernova remnant.
It is a well-known result that there are three stages in the evolution of
a supernova remnant in the interstellar medium (Spitzer 1978).
First, the remnant expands freely at nearly constant speed into the
surrounding medium.
In this phase, the deceleration due to the swept-up interstellar matter in 
front of the shock does not intervene in the expansion.
As the shock plows into the interstellar matter, it begins to
decelerate.
When the mass of the swept-up matter $(4\pi/3)R_s^3\rho_{\rm ISM}$ becomes
nearly equal to 
the mass of the ejecta $M_{\rm ejecta}$, one needs to take into account 
the deceleration of the shock front.
Thereafter the evolution goes into the next 
stage, the Sedov (adiabatic) phase.
In this Sedov phase, the temperature behind the shock front is so high that 
deceleration due to cooling behind the shock front can be ignored
compared with the expansion.
There is a well-known similarity solution for the
evolution of the shock front,

\begin{eqnarray}
R_s &=& \xi_0(Et^2/\rho_{\rm ISM})^{1/5}, \\
v_s &=& 0.4\xi_0(E/\rho_{\rm ISM}t^3)^{1/5},
\end{eqnarray}
where $E$ is the initial energy input and $\xi_0=1.17$ with
$\gamma$=5/3 (Shu 1992).
As the shock decelerates and the temperature behind the front decreases,
cooling begins to affect the expansion and a dense shell is formed
behind the shock front.
Almost all the mass of the ejecta and plowed-up interstellar medium are in 
the shell, and the thermal 
energy of the remnant is radiated away by the shell.
The condition to be satisfied at the transition from the
Sedov phase to the radiative phase is $\tau_{\rm ex}\equiv
R_s/v_s\approx\tau_{\rm cool}|_{\rm shock}$. 
When this condition is satisfied, i.e., the cooling time subsequently 
becomes shorter than
expansion time of the shell, the thermal energy of the remnant is
radiated very rapidly (Chevalier 1974).

We will examine the evolution of the cocoon after the jet turns
off by analogy with the evolution of the supernova remnant described above.
Cocoons should have three evolutionary stages as do supernova remnants.
Let the cocoon evolve into the Sedov phase at $t=t_s$ and into the
radiative phase at $t=t_r$ with $t$=0 at the time when the jet stops.
First, the cocoon expands with nearly constant speed.
The shock surrounding the cocoon engulfs the intergalactic medium.
Accordingly the kinetic energy is turned into the thermal energy, and  
the fraction of the thermal energy of the 
shock-compressed IGM to the total energy input by the jet,  
$U_{\rm IGM}(t)/(L_jt_{\rm life})$, increases with time.
Let us consider similarly as the case of supernova remnant further.
The deceleration by the piled-up IGM cannot be ignored when 
the fraction of kinetic energy, $|\Delta E_k/E_k|$, 
which is transformed into thermal energy, 
approaches unity.
At this epoch we can consider that 
the Sedov phase sets in.
As we wrote in \S3, while the jet is on, $U_{\rm IGM}$ is the same order
as $E_k$, and therefore one can regard that 
the Sedov phase begins when $|\Delta E_k(t_s)/E_k(0)|\sim
\Delta U_{\rm IGM}(t_s)/U_{\rm IGM}(0)={\cal O}(1)$.
In other words, $t_s$ is about the same as the time required to
double the thermal energy of the cocoon at the time when the jet stops
($t=0$).
Remembering that the cocoon expands almost constantly before the Sedov
phase because of lack of swept material which decelerates the
expansion, i.e., $\Delta U_{\rm IGM}(t)/\Delta t \sim L_j \sim U_{\rm
  IGM}(0)/t_{\rm life}$,  
we can consider that the cocoon evolves into the Sedov phase at
$t=t_s\approx t_{\rm life}$, since $\Delta U_{\rm IGM}(t_s)\sim U_{\rm
  IGM}(0)$. 

The radiative phase begins when the cooling time behind the shock front
$\tau_{\rm cool}|_{\rm shock}$ 
catches up with the expansion time of the Sedov phase
$R_s/v_s$.
Because the thermal energy of the cocoon is radiated away very rapidly once 
cooling sets in, we expect that the cocoon remains hot enough to be the
source of the Sunyaev-Zel'dovich effect only during the first two
phases.
Therefore in effect the lifetime of the cocoon can be taken as $\sim t_r$.

In order to know $t_r$, we need to evaluate $\tau_{\rm cool}|_{\rm
  shock}$ 
and hence 
we need to know the cooling processes of the cocoon.
First we examine the content of the cocoon after the jet is off.
The total thermal energy of the cocoon is the sum of that
of the jet-supplied matter and of the shock-compressed
intergalactic matter.
Within the model described in the previous sections, the jet-supplied matter
and the compressed intergalactic matter are in pressure equilibrium,
and we therefore cannot distinguish the former from the latter.
Thus we do not know the ratio of the thermal energy of the jet-supplied 
matter to the total energy.
Hereafter we assume that the thermal energy of the shock-compressed
matter already dominates that of the jet-supplied matter at the time
when the jet stops.

We need to include various cooling processes that affect the temperature 
behind the shock front,
\begin{equation}
T_c=\frac{3v_c^2\mu}{16k}. \label{Tc}
\end{equation}
From the discussion of 
synchrotron spectral aging, the shock speed $v_c$ is estimated to
be $v_c<v_h\lesssim 0.1c$ (Liu et al. 1992).
This result means that $T_c$ is lower than $\sim 10^{10}$ K from
equation (\ref{Tc}).
We can thus ignore pair creation ($e^--e^+$) within the cocoon.
We regard the cocoon matter as a hydrogen plasma ($p+e^-$).
The dominant cooling processes within this plasma are thermal bremsstrahlung 
and inverse Compton cooling, and 
hereafter we take account of these processes for the
estimate of $\tau_{\rm cool}|_{\rm shock}$.

\begin{figure}[t]
\centerline{\epsfxsize=10cm \epsffile{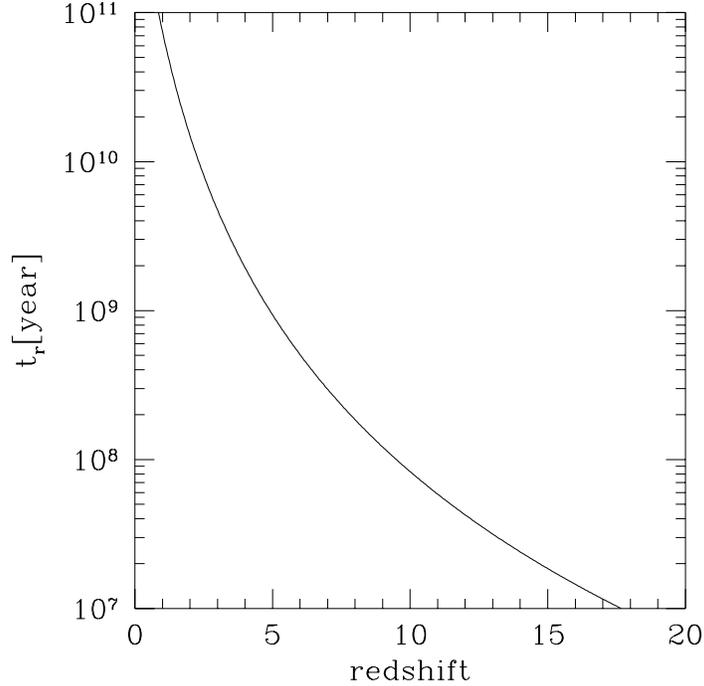}}
\caption{\protect\footnotesize
Effective lifetime of the cocoon after the jet is 
  off $t_r$ against redshift. At higher redshift Compton cooling is more 
  effective and $t_r$ is shorter. 
}
\label{tr}
\end{figure}
In figure \ref{tr}, we plot $t_r$ versus redshift.
At higher redshift, inverse Compton cooling is more efficient and the 
cooling time
is shorter.
As noted above, $t_s$ is about the same order as the jet lifetime
$t_{\rm life}\approx 10^7-10^8$ years.
This means that when $t_r<t_{\rm life}$, $t_r<t_s$ and no Sedov phase 
would exist.
In this case we set $t_r=10^6$ years (the final result is hardly
affected by this value).
On the other hand, when $t_r>t_{\rm life}$, $t_r>t_s$ and the Sedov
phase exists. 
We make a fitting formula for $t_r$ as a function of redshift:

\begin{eqnarray*}
0<z<5 &:& \log_{10}(t_r/{\rm years})=10.7-0.37z ~ , \\
5<z<10 &:& \log_{10}(t_r/{\rm years})=9.93-0.21z ~ , \\ 
10<z<20 &:& \log_{10}(t_r/{\rm years})=8.96-0.11z ~ . 
\end{eqnarray*}
The thermal energy of the cocoon is expected to decrease rapidly at
$t>t_r$ so we assume 

\begin{equation}
P_cV\propto \exp\left( -\frac{t}{t_r} \right), \label{ff} 
\end{equation}
after the jet stops.

\section{Averaged $y$ Estimate}
In the previous sections, we have obtained all the building blocks for a single
cocoon in order to calculate the total $y$.
One remaining parameter is the cocoon number density $n(M,z)$.
We employ the Press-Schechter formalism, assuming $n(M,z)=f_r(M,z)\times 
n_{\rm PS}(M,z)$, where $f_r(M,z)$ is the fraction of radio galaxies among
the objects which have mass $M$ at $z$.
In the next subsection we will model $f_r(M)$.

\subsection{Estimation of $f_r$}
As we described in \S4, cocoons can remain hot for a long time.
Therefore in order to calculate $y$, we should count the
``relics'' of radio
galaxies.
However, we do not have any theoretical model of $f_r(M)$: instead we
make a rough estimate of $f_r(M)$ from the present-day luminosity functions
of radio galaxies and field galaxies.\footnote{In this subsection we 
  use $h=0.5$ and $q_0=0.0$ in order to use the results in the
  literature.} 

First, for radio galaxies, we make use of the local luminosity function
of radio sources of Condon (1989).
His sample contains both the active galaxies with massive black hole
(``monsters'') and starburst galaxies.
But these two types of galaxies cover quite a different range of radio power: 
most of the extragalactic radio sources with $P(1.4{\rm GHz})<10^{23}$W
Hz$^{-1}$ are starburst galaxies and most of the sources with
$P(1.4{\rm GHz})>10^{23}$W Hz$^{-1}$ are ``monsters''(see his figure 9;
see also Sopp \& Alexander 1991).

\begin{figure}[t]
\centerline{\epsfxsize=10cm \epsffile{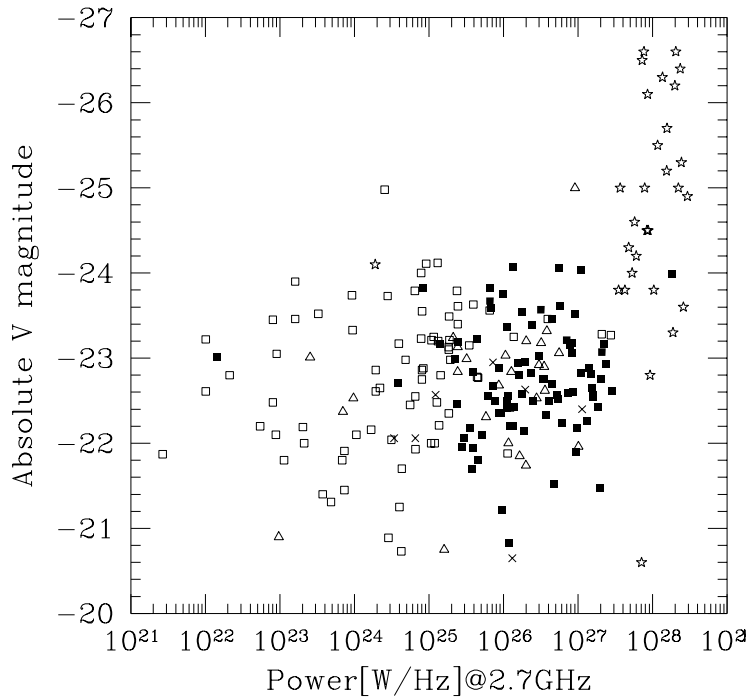}}
\caption{\protect\footnotesize
Radio power at 2.7 GHz and absolute $V$ magnitude
  relation  for radio galaxies. Data is taken from the table of Zirbel \& Baum
  (1995). Open squares are FRI galaxies, filled squares are FRII
  sources, crosses are intermediate ones, triangles are those with
  unidentified morphology, and open stars are high $z(>0.5)$ galaxies. 
  Zirbel \& Baum (1995) adopted $h=0.5$ and $q_0=0.0$. }
\label{vradio}
\end{figure}

\begin{figure}[t]
\centerline{\epsfxsize=10cm \epsffile{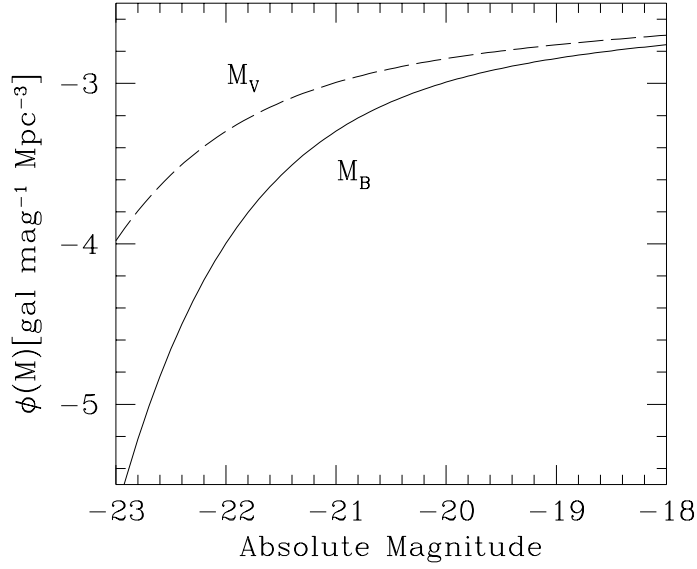}}
\caption{\protect\footnotesize
Schechter function in terms of $M_V$ and $M_B$.  
}
\label{vshechter}
\end{figure}

Next, in order to compare the numbers of radio galaxies with ordinary
galaxies, we examine the (optical luminosity)-(radio power)
relation for radio galaxies.
We employ the data of Zirbel \& Baum (1995) and plot the
$P(2.7{\rm GHz})$-absolute $V$ magnitude relation.\footnote{We make use
  of the power law index $\alpha$ in the Table of Zirbel \& Baum (1995) 
  to calculate $P(2.7{\rm GHz})$ from $P(408{\rm MHz})$, and when
  $\alpha$ is not displayed in the Table, we set $\alpha=0.75$.}
From figure \ref{vradio}, we can see that there is an upper limit 
of absolute magnitude $M_V\sim -$22 mag though there is a large scatter.
This means that there is a small number of radio galaxies fainter than
$-$22 mag.
Finally, for the luminosity function of field galaxies we adopt the
Schechter function.
Note that usually the Schechter function is a function of $M_B$, not of $M_V$.
Therefore we have to obtain the $M_V$-Schechter function.
We include 4 Hubble types of galaxies, E/S0, S$_{\rm ab}$, S$_{\rm
  cd}$,and 
S$_{\rm dm}$, and adopt the luminosity-averaged local fraction of each types 
from Yosii \& Takahara (1988).
We adopt the value $\psi^{\ast}=2.3\times 10^{-3}$ Mpc$^{-3}$,
$M_B^{\ast}=-21.1$, and $\alpha=-1.11$ from Yosii \& Takahara (1988),
and the averaged $\bv$ colors for each morphological type of galaxies 
are taken from
Babul \& Ferguson (1996).
The resulting $M_V$-Schechter function is displayed in figure
\ref{vshechter}.
Therefore we obtain a model of the fraction of radio galaxies in terms of

luminosity,

\begin{eqnarray}
f_r(M_V<-22 {\rm mag}) &\approx& \frac{\int \phi(P_{\rm radio})dP_{\rm
    radio}}{\int^{-22}\phi_{\rm Schechter}(M_V)dM_V} \approx 0.01, \nonumber \\
f_r(M_V>-22 {\rm mag}) &\approx& 0,
\end{eqnarray}
where $\phi$ is a luminosity function of radio galaxies and $\phi_{\rm
  Schechter}$ is the $M_V$-Schechter function.

Because we use the mass function of galaxies to count the cocoons, we
need $f_r$ as a function of mass.
From our observation of figure \ref{vradio} that radio galaxies have a
lower bound to their optical luminosity, 
we further assume that radio galaxies have a
lower mass limit corresponding to  $M_V\sim -$22 mag. 
The critical magnitude $M_V=-22$ mag is about the same as $L^{\ast}_V$
($\bv \approx 1.0$), and the lower mass limit to the total (baryons+dark
matter) mass of a radio galaxy can be taken to be $\sim 10^{12}M_{\sun}$.
It is natural to consider that $f_r$ depends only on masses greater than
$10^{12}M_{\sun}$ but moreover we
assume that for simplicity $f_r$ is independent of $M$ in this mass range.
Then we can model $f_r(M)$ as

\begin{eqnarray}
f_r(M) \equiv f_r^{\ast} = f_r(M_V<-22 {\rm mag})=0.01 & & M>10^{12}M_{\sun},  \nonumber \\
f_r(M) = 0               & &           M<10^{12}M_{\sun}.
\end{eqnarray}

From the popular idea that the host galaxy of a radio galaxy is a giant
elliptical, we set the upper limit on the mass to be $10^{14}M_{\sun}$.

\subsection{Integration and Results}
The physical number density of cocoons is proportional to 

\begin{equation}
n_{\rm PS}(M,z)dM = \sqrt{\frac{2}{\pi}}\frac{\rho(z)}{M}\frac{\delta_c}{D(z)} 
  \left[ -\frac{1}{\sigma^2}\frac{\partial\sigma}{\partial M}\right] 
\exp\left[ -\frac{\delta_c^2}{2\sigma^2D(z)^2}\right]dM,
\end{equation} 
where $\rho(z)$ is the proper density at $z$, 
$\delta_c=1.69$, $D(z)$ is the growth of the perturbation
($D(z)=(1+z)^{-1}$ for the Einstein-de Sitter universe), and $\sigma^2$
is the variance with a Gaussian filter,

\begin{equation}
\sigma^2(M)=\frac{1}{2\pi^2}\int^{\infty}_{0}P(k)\exp(-r_M^2k^2)k^2{\rm
  d}k.
\end{equation}
Here $r_M=(M/\rho_0)^{1/3}/\sqrt{2\pi}$ is the smoothing radius of the
Gaussian filter with present critical density $\rho_0$, and $P(k)$ is
the power spectrum of density perturbations.
In calculating $P(k)$ we use the fitting formula of Bardeen et
al. (1986),

\begin{eqnarray}
P(k) &=& AT(k)^2k, \\
T(k) &=&
\frac{\ln(1+2.34q)}{2.34q}[1+3.89q+(16.1q)^2+(5.46q)^3+(6.71q)^4]^{-1/4}, 
\\
q &\equiv& \frac{k}{\Omega h^2{\rm Mpc}^{-1}}. 
\end{eqnarray}
In this paper we set $\Omega=1.0$ and $h=0.8$.
For the normalization factor $A$ we use the 
{\it COBE}-normalized value (Sugiyama 1995) and approximate 
\begin{equation}
-\frac{1}{\sigma^2}\frac{\partial\sigma}{\partial M}\approx
\frac{1}{M}\left( \frac{(n+3)}{6}\right), 
\end{equation}
with $P(k)\propto k^n$ within the range of mass of a radio galaxy.
With $10^{12}M_{\sun}\le M \le 10^{14}M_{\sun}$, $-2.5\lesssim n
\lesssim 
-0.5$ so
we set $n=-2$ throughout.

Thus the total averaged $y$ is given by 
\begin{eqnarray}
y &=& \int \left( \int_{\rm single ~ cocoon}\frac{kT}{m_ec^2}n\sigma_TdR
\right) f_rn_{\rm PS}(M,z)\frac{\theta^2(M,z)}{d\Omega} a^3r^2drdMd\Omega,
\nonumber \\
&\approx& \int
\frac{P_cR}{m_ec^2}\sigma_Tf_rn_{\rm PS}(M,z)\theta^2(M,z)
a^3r^2drdM.
\end{eqnarray}

Therefore, 
\begin{eqnarray}
y &=& \int f_r
\frac{P_cR^3}{m_ec^2}\sigma_Tn_{\rm PS}(M,z)\frac{1}{R_A^2}a^3r^2drdM,
\nonumber \\
&=& \int f_r
\frac{P_cV}{m_ec^2}\sigma_Tn_{\rm PS}\frac{1}{R_A^2}\frac{27}{2}(ct_0)^3 
(1+z)^{-9/2}\left( 1-\frac{1}{\sqrt{1+z}}\right) ^2dzdM. \label{ytot} 
\end{eqnarray}
If we use equations (\ref{Pc}) and (\ref{V}),
\begin{equation}
 P_cV= \cases{
      L_j(\gamma-1)t, & :$t<t_{\rm life}$ \cr
      L_j(\gamma-1)t_{\rm life}\times\exp(-(t-t_{\rm life})/t_r(z)), &
      :$t>t_{\rm life}$ \cr
}  \label{tot}
\end{equation} 
applying equation (\ref{ff}).

We integrate equation (\ref{ytot}) with $10^{12}M_{\sun}\le M \le
10^{14}M_{\sun}$ and $0\le z\le 20$.
For the lifetime of the jet we take $t_{\rm life}=10^7$ years regardless 
of mass.
We obtain, 

\begin{equation}
y_{\rm tot}\approx 5.72\times 10^{-5}\left
  ( \frac{f_{\rm BH}}{0.002}\right) \left( \frac{f_r^{\ast}}{10^{-2}}\right). 
\label{result}
\end{equation}
As we discussed in \S4, cocoons serve as the source of
the Sunyaev-Zel'dovich effect after the jet is off, and the number of the
cocoons can be greater than the number of the luminous radio galaxies.

Our estimate of $y$ should be compared with the {\it COBE} FIRAS 
constraint $y\le 1.5\times 10^{-5}$ (Fixen et al. 1996).
Discussions about our result and this constraint are presented in the
final section.
\clearpage
\section{Discussion}
We have investigated the Sunyaev-Zel'dovich effect produced 
by cocoons of radio
galaxies.
We have employed a simple model of Nath (1995) while the jet is on.
We have also examined 
the evolution of a cocoon after the jet turns off, adopting the analogy to the 
evolution of a supernova remnant in the interstellar medium.
The cocoon remains at high temperature after the jet turns off 
for over $\gtrsim 10^8$ years to low
redshift $z\lesssim 10$, remaining overpressured against the pressure of the
IGM.
Because of the short lifetime of the jet ($\sim 10^7-10^8$ years), the
main contribution to $y$ comes from the cocoons  
with the jet off. 
Comparing our result with the {\it COBE} constraint we 
find that hot cocoons could be significant foreground 
sources of the CMB.
We discuss the assumptions and the implications  
of the models below.

We have made a model for $f_r(M)$ which is a step function with a
cut-off on the upper end of the mass range.
We assumed the lower limit on the mass of a radio galaxy to be
$10^{12}M_{\sun}$ from the optical magnitude distribution for radio
galaxies, but this value has some ambiguity.
We performed calculations changing the lower mass limit $M_{\rm low}$,
and found that $y$ increases very rapidly with decreasing $M_{\rm low}$
to exceed the ${\it COBE}$ constraint by far.
This might support the view that there exists a low mass limit for jet 
activity.

\begin{figure}[t]
\centerline{\epsfxsize=10cm \epsffile{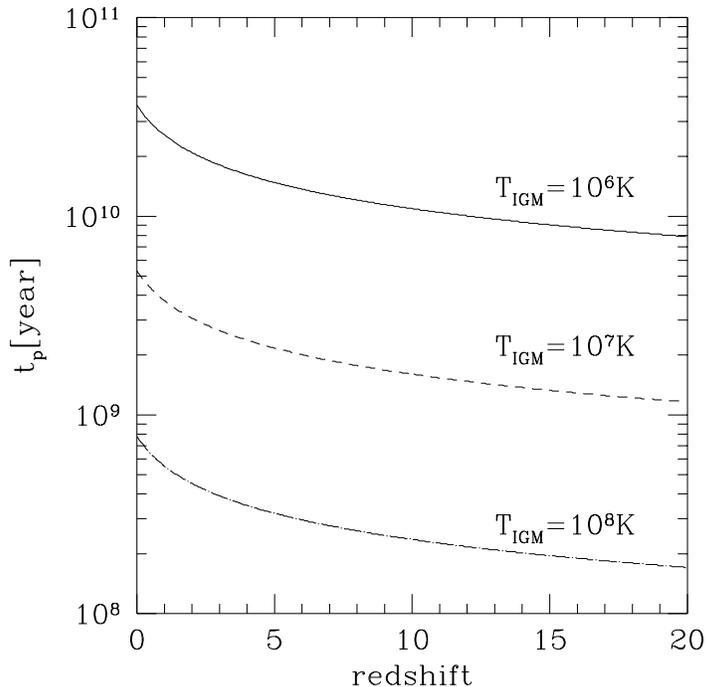}}
\caption{\protect\footnotesize
Timescale to reach the state of 
pressure equilibrium with the
  external IGM pressure $t_p$ against redshift. The temperature of the
  IGM is constant. Solid line corresponds to the case with $T_{\rm
  IGM}=10^6$ K and dashed line corresponds to the case with $T_{\rm IGM} 
  =10^7$ K.}
\label{tp}
\end{figure}

In \S4 we have presumed that the cocoon remains overpressured against 
the intergalactic medium.
However, we must investigate the possibility that the pressure of the
cocoon decreases to be in pressure equilibrium with the surrounding
medium within the effective lifetime of the cocoon $\sim t_r$.
Once in pressure balance with the IGM, the internal energy of the cocoon 
is consumed by $PdV$ work.
We calculate the timescale to be in pressure equilibrium with IGM $t_p$ as a
function of redshift.
There has not been any direct observational means for obtaining the 
pressure of the IGM.
As for the temperature $T_a$, one can expect $T_a\gtrsim 10^4-10^6$ K
from the previous studies of the reionization of the Universe
(see e.g., Fukugita \& Kawasaki 1994;Tegmark, Silk, \& Blanchard 1994).
However, radio galaxies or radio loud quasars are likely to be found in
high density regions (Jones et al. 1997).
Pressure in such an environment is higher than in an ordinary region.
Thus  we take the temperature of the intergalactic
medium $T_{\rm IGM}$ to be $10^6-10^8$ K, on the order of the
present value of the temperature of the intracluster gas.
For the number density of IGM $n_a$, as an extreme case, we equate $n_a$ 
with the critical density $n_{\rm cr}=3H_0^2/(8\pi Gm_p)(1+z)^3$, which
is much larger than the average baryon density by a factor of $\Omega_bh^2$.
On the other hand, we make use of the similarity solution for the Sedov
phase to calculate the pressure behind the shock front.
In figure \ref{tp} we plot $t_p$ against redshift $z$.
From this figure we see that the lifetime of a cocoon in a rich cluster
of galaxy is shorter than $t_r$.
For the most extreme case, with the electron number density $n_e\sim
10^{-3}$cm$^{-3}$ and the temperature $T\sim 10^8$K, $t_p$ is about
$1\times 10^8$years. 
We find that when $t_r\ge 10^8$years for $z\le 10$, $y\gtrsim
2.6\times 10^{-6}(f_{\rm BH}/0.002)(f_r/10^{-2})$, which is still
observable.

For radio luminous galaxies, the radio emission itself may affect 
the Sunyaev-Zel'dovich decrement.
In order to estimate this effect 
we should take into account the integrated spectrum in a line of sight
from cocoons at different redshift.
From equations (17) and (33), however, we see that there is marginally one
cocoon in one line of sight, which means that the contamination by radio
emission from other radio galaxies is not a severe problem.
Thus we only 
have to consider the radio emission accompanying the cocoon itself.
Detailed spectrum evolution requires a specific model of radio lobes, though,  
we just present simple discussions in this paper.

The Sunyaev-Zel'dovich 
decrement of flux is (Zel'dovich \& Sunyaev, 1969)
\begin{equation}
 \frac{\Delta I_{\nu}}{I_{\nu}}=\frac{xe^x}{(e^x-1)}
 \left[ x\left(\frac{e^x+1}{e^x-1}\right)-4\right] \cdot y,
\end{equation}
with the nondimensional frequency $x\equiv h\nu/kT$. 
In the Rayleigh-Jeans region($x\ll 1$),
\begin{equation}
 \frac{\Delta I_{\nu}}{I_{\nu}}\to -2y,
\end{equation}
which is independent of $x$, and the CMB flux containing
Sunyaev-Zel'dovich signal has also black body spectrum.
At $\nu\gtrsim 10$GHz, the spectrum of radio galaxy is
$I_{\nu}\propto \nu^{-\alpha} ($typically $\alpha >0$ for bright
sources; see Sokasian et al. 1998), which is quite
different from the CMB spectrum ($\propto \nu^2$).
As Sokasian et al. (1998) pointed out, bight radio galaxies can be major 
sources of the anisotropy of CMB especially on subdegree scales at
10GHz$\lesssim\nu\lesssim$200GHz, and the Sunyaev-Zel'dovich singal can be
completely overcome by the radio emission in this waveband and angular
scale while lobes
are bright.
In the Wien region, radio emission from radio galaxies becomes
increasingly fainter with $\nu$, 
and the influence on the deformation of CMB is far 
weaker than in the Rayleigh-Jeans region.
Thus, one coule consider that 
the observational prediction of the integrated $y$ 
of the Sunyaev-Zel'dovich effect 
is significantly contaminated by emission from the lobes in low
frequencies. 
In such a case, the multiwavelength observation covering a wade
range including both Rayleigh-Jeans and Wien region, which PLANCK
satellite will carry out\footnote{see 
  http://astro.estec.esa.nl/SA-general/Projects/Planck/.}, 
 is required in
order to detect the Sunyaev-Zel'dovich effect. 

However, the effect of radio emission from the lobes depends on the 
ratio of the timescale of energy loss of the high energy electrons by
synchrotron 
emission and the cocoon effective lifetime($\approx t_{\rm life}+t_r$).
Synchrotron emission from lobes may not fade away immediately after
the jet stops, while cocoons can survive for a relatively long period.
If the lifetime of synchrotron
emission is significantly shorter than the effective cocoon lifetime, we 
expect little contamination from radio emission to measure the
Sunyaev-Zel'dovich signal in the Rayleigh-Jeans region.  
The loss timescale can be estimated if we know the magnetic field
strength and the density of high energy electrons at the lobes.
Note that McKinnon, Owen, \& Eilek (1991) reported the detection of
Sunyaev-Zel'dovich decrement from radio jet lobes with NRAO 12m
telescope.

The lobe decay time is also closely related to the fraction of high energy
electrons $\epsilon_t$ (see below).
However, the estimate of
the lobe decay time requires a detailed model of radio emitting lobes, 
and this 
is beyond the scope of this paper.
In addition
to the evolution of the spectrum, the estimate of lobe emission 
decay time should be
regarded as a future work.

Radio galaxies have one or two lobes that emit synchrotron radiation 
from high energy electrons, 
which are accelerated via  
Fermi acceleration at the shock.
We assumed in the cocoon model that almost all of the jet matter is
    completely thermalized at the shock, but more realistically
    speaking, only a fraction $\epsilon_t<1$ of the jet-supplied energy 
can become the thermal
    pressure of the cocoon.
Therefore $L_j$ in equation (\ref{tot}) is replaced by $\epsilon_t L_j$
and $y$ is $\epsilon_ty$.
As expressed in \S 3, when the jet is stopped by the shock, 
a large fraction of jet matter would be thermalized
($\epsilon_t\approx 1$) with the magnetic field.
The thermalization efficiency 
$\epsilon_t$ can be estimated if we know the species of the jet particles
(whether they are neutral particles or electron-positron plasma as
expected by the central engine of a massive black hole) and the
acceleration mechanism at the shock.
X-ray emission from the lobes, which is considered to be inverse
Compton scattered synchrotron photons, will provide important information
about $\epsilon_t$.

Jones et al. (1997) found the decrement of CMB
temperature towards the $z=3.8$ radio-quiet quasar pair PC1643+4631 A
and B.
There is no detected optical or X-ray source in this direction, and 
they consider that a low surface brightness cluster of
galaxies may be present at relatively high ($1\le z \le 3.8$) redshift.
Another explanation is suggested by Natarajan \& Sigurdsson
(1999), who consider that the decrement is the kinematical
Sunyaev-Zel'dovich effect produced by quasar winds.
Neither explanations seems particularly plausible.
We suggest that the so-called ``black cloud'' phenomenon 
(see also Richards et al. 1997) may be related to  
the relics of radio galaxies, with the lobes
already vanished and with the cocoon still hot enough to give a sizable 
Sunyaev-Zel'dovich decrement.

\acknowledgements
We thank H. Susa and R. Nishi for useful discussions in the early stage 
of this study, and anonymous referee for helpful comments.
M.Y. is grateful for many valuable comments by K. Ohta about the
luminosity functions.
The research of J.S. has been supported in part by a grant from NASA. 
This work is financially supported in part by 
the Grant-in-Aid for Encouragement of Young Scientists 
by the Ministry of Education,
Culture, Science and Sports, No.09440106.


\end{document}